\documentclass[submission,copyright,creativecommons]{eptcs}
\providecommand{\event}{} % Name of the event you are submitting to
\usepackage{amsmath}
\usepackage{amssymb}
\usepackage{xcolor}
\newcommand{\mkwd}[1]{\mathsf{#1}}
\newcommand{\globalsend}[4]{\role{#1} \rightarrow \role{#2} : \mkwd{#3}(#4)}
\newcommand{\globalsendann}[5]{\role{#1} \rightarrow \role{#2} : \mkwd{#3}\{#4\}(#5)}

\newcommand{\globalend}{\mkwd{end}}
\newcommand{\localend}{\mkwd{end}}

\newcommand{\globalbranch}[2]{\role{#1} \rightarrow \role{#2} \: : \{}
\newcommand{\globalmsg}[2]{\mkwd{#1}(#2)}
\newcommand{\globalmsgann}[3]{\mkwd{#1}\{#2\}(#3)}
\newcommand{\role}[1]{{\color{purple}{\ensuremath{\mathttbf{#1}}}}}
\DeclareMathAlphabet{\mathttbf}{\encodingdefault}{\ttdefault}{bx}{n}
\newcommand{\var}[1]{\mathit{#1}}
\newcommand{\ty}[1]{\mkwd{#1}}
\newcommand{\then}{\:.\:}

\newcommand{\msg}[2]{#1(#2)}
\newcommand{\msgtag}[1]{\mkwd{#1}}

\newcommand{\localselectone}[1]{\role{#1} \mathop{\oplus} \{}

\newcommand{\localoffersingle}[3]{\role{#1} \mathop{\&} \msg{#2}{#3}}

\newcommand{\qqquad}{\qquad \quad}
\usepackage{graphicx}

\usepackage{iftex}
\usepackage{microtype}
\usepackage[sort&compress,numbers]{natbib}
\usepackage{orcidlink}

\ifpdf
  \usepackage{underscore}         % Only needed if you use pdflatex.
  \usepackage[T1]{fontenc}        % Recommended with pdflatex
\else
  \usepackage{breakurl}           % Not needed if you use pdflatex only.
\fi

\title{Behavioural Types for Heterogeneous Systems \\ (Position Paper)}
\author{
Simon Fowler
\institute{University of Glasgow, UK}
\and Philipp Haller
\institute{Digital Futures, KTH Royal Institute of Technology, SE}
\and Roland Kuhn
\institute{Actyx AG, DE}
\and Sam Lindley
\institute{The University of Edinburgh, UK}
\and Alceste Scalas
\institute{Technical University of Denmark, DK}
\and Vasco T.\ Vasconcelos
\institute{University of Lisbon, PT}
}
\def\titlerunning{Behavioural Types for Heterogeneous Systems (Position Paper)}
\def\authorrunning{Fowler \emph{et al}.}
\begin{document}
\maketitle

\begin{abstract}
Behavioural types provide a promising way to achieve lightweight,
language-integrated verification for communication-centric software. However, a large barrier to the adoption of behavioural types is that the current state of the art expects software to be written using \emph{the same} tools and typing discipline throughout a system, and has little support for components over which a developer has no control.

This position paper describes the outcomes of a working group discussion at Dagstuhl Seminar 24051 (Next-Generation Protocols for Heterogeneous Systems). We propose a methodology for integrating multiple behaviourally-typed components, written in different languages.
Our proposed approach involves an \emph{extensible protocol description language}, a \emph{session IR} that can describe data transformations and boundary monitoring and which can be compiled into program-specific \emph{session proxies}, and finally a \emph{session middleware} to aid session establishment.

We hope that this position paper will stimulate discussion on one of the most pressing challenges facing the widespread adoption of behavioural typing.
\end{abstract}

\section{Introduction}
Behavioural types provide a powerful and lightweight mechanism for language-integrated verification of behavioural properties: whereas traditional data types rule out errors such as adding an integer to a string, behavioural types can rule out behavioural errors such as forgetting to close a file handle or sending an invalid message on a communication channel. 

\emph{Session types}~\cite{Honda93,HondaVK98} are a behavioural typing discipline for checking adherence to communication protocols: if a process is typed according to its session type, then it is guaranteed to fulfil its role in the communication protocol at runtime. Although originally designed for two communicating participants, work on \emph{multiparty} session types (MPSTs)~\cite{HondaYC08} extends session typing to handle systems with multiple components. 
If all components are either derived from a well-formed global type, or all components have compatible local types, then the entire system should not encounter any communication errors at runtime. Many MPST disciplines include further guarantees such as global progress or liveness.

Extensive research has gone into behavioural types, in particular session types, over the years. A recent workshop celebrated 30 years of session types~\cite{st30}, and a recent book concentrates on how decades of research into behavioural types has given rise to a plethora of tools~\cite{GayR17}. Many international programming languages conferences typically have sessions dedicated to behavioural type systems.

Nevertheless, and notwithstanding efforts to overcome barriers to practical adoption (e.g., work on failure handling~\cite{BarwellSY022,FowlerLMD19}, graphical user interfaces~\cite{MiuFYZ21,Fowler20}, and subsessions~\cite{DemangeonH12}), behavioural typing has not yet seen widespread industrial use. Arguably \emph{the} key issue facing the behavioural types community\footnote{And indeed, other neighbouring communities such as choreographic programming~\cite{Montesi23}, although we concentrate on behavioural types in this paper.}
is the inability for behavioural types to work satisfactorily with \emph{heterogeneous} systems, consisting of software components developed in different languages, using different tools, with different typing guarantees. 

Heterogeneity may arise for practical reasons: for example, we may want to write some components in a given language due to better library support (e.g., using Python due to its rich support for data science), or because it is important to obtain stronger static guarantees for a particular participant or portion of a protocol.
Heterogeneity may even arise in a single program, for example writing parts of
an application in different programming languages (e.g., writing
performance-critical code in a systems language such as C++ or Rust, and the
remainder of the program in a managed language like Python).
Furthermore, in any realistic setting, we would have to assume
that some components are implemented using different languages or accessible
only through an API (for example as is common with microservices).
Similarly, we may have existing components that offer \emph{similar} services
that do not quite correspond to the expected types.
We concentrate on the following scenario:

\vspace{-3mm}%
\begin{quote}
\textbf{Scenario: Travel Booking System.} We want to design a travel booking system that includes both timing constraints and data refinements.
A travel booking session consists of a client, travel agent, and flight
provider; we are in control of the agent and client, but the provider is developed by an external company and accessible via an API.
The client initiates a search, and receives a set of suitable flights.
After receiving the results, the client has 6 minutes to select a flight before the results expire.

When the client selects a flight, it receives a token, and should send the same token to the provider in order to continue the booking.
Finally, the provider sends the client either a booking confirmation, or an error message.
\end{quote}
Crucially, we will consider a heterogeneous version of this scenario
in which the client, travel agent, and flight provider are implemented
using different tools with quite different capabilities.

\paragraph{Paper structure.}
The paper proceeds as follows.
Section~\ref{sec:background} gives relevant background.
Section~\ref{sec:hetero-mpst} describes a motivating scenario of a travel agent application written in different tools with differing language features. Section~\ref{sec:solution} describes our proposed solution and speculates on several potential research challenges.
Section~\ref{sec:related} discusses related work, and Section~\ref{sec:conclusion} concludes.

\section{Background}\label{sec:background}

\paragraph{(Multiparty) Session Types.}
Session types are types for protocols: whereas a data type describes the shape of some data, ruling out errors such as adding an integer to a string, a session type describes both the type and direction of data to be communicated between participants~\cite{Honda93, HondaVK98}. Session types were originally investigated in the \emph{binary} setting between two participants, but later work on \emph{multiparty} session types~\cite{HondaYC08} describes communication between multiple communicating participants. We concentrate on multiparty session typing in the remainder of the paper.

The following example describes the classic \emph{two-buyer} protocol where two participants collaborate in order to buy an expensive item (usually a book). \role{Buyer1} begins by sending the title to the \role{Seller}, who responds with a quote. \role{Buyer1} then sends the quote to \role{Buyer2}, who decides whether to accept the quote by sending their address to the \role{Seller} and subsequently receiving a delivery data, or declining the offer.

The \emph{global type} describing all interactions in the system is described on the left. Global types can then be \emph{projected} into \emph{local types} for each participant; it is then possible to typecheck or monitor all participants against their local types. The local type for \role{Buyer2} is shown on the right.

\begin{center}
\begin{minipage}{0.4\columnwidth}
{\footnotesize
\[
    \hspace{-4.5em}
    \begin{array}{l}
    \globalsend{Buyer1}{Seller}{\msgtag{title}}{\ty{String}} \then \\
    \globalsend{Seller}{Buyer1}{\msgtag{quote}}{\ty{Int}} \then \\
    \globalsend{Buyer1}{Buyer2}{\msgtag{share}}{\ty{Int}} \then \\
    \globalbranch{Buyer2}{Seller} \\
    \quad \msg{\msgtag{address}}{\ty{String}} \then \\
    \qquad \globalsend{Seller}{Buyer2}{\msgtag{date}}{\ty{Date}}
    \then \globalend, \\
    \quad \msg{\msgtag{quit}}{\ty{Unit}} \then \globalend \\
    \}
    \end{array}
\]
}
\end{minipage}
\qquad
\begin{minipage}{0.4\columnwidth}
{\footnotesize
\[
    \begin{array}{l}
    \mathsf{Buyer2} \;\triangleq\; \localoffersingle{Buyer1}{\msgtag{share}}{\ty{Int}} \then \\
    \phantom{\mathsf{Buyer2} \;\triangleq\;}\;\,
    \localselectone{Seller} \\
                     \phantom{\mathsf{Buyer2} \;\triangleq\;}\qquad \msg{\msgtag{address}}{\ty{String}} \then \\
                     \phantom{\mathsf{Buyer2} \;\triangleq\;}\qqquad
                       \localoffersingle{Seller}{\msgtag{date}}{\ty{Date}}
                       \then \localend, \\
                     \phantom{\mathsf{Buyer2} \;\triangleq\;}\qquad \msg{\msgtag{quit}}{\ty{Unit}} \then
                \localend \\
    \phantom{\mathsf{Buyer2} \;\triangleq\;}\;\, \}
    \vspace{0.5em}\\
    \end{array}
\]
}
\end{minipage}
\end{center}

A multitude of tools have been developed for checking against multiparty session types, for example in Java~\cite{KouzapasDPG18}, Scala~\cite{DBLP:conf/ecoop/ScalasDHY17}, Rust~\cite{LagaillardieNY22,CutnerYV22}, and F\#~\cite{NeykovaHYA18}. However, each tool embeds the assumption that the entire system is written using that same tool; the possibility of combining heterogeneous components written using different tools and programming languages is not part of the tools' specification.

\paragraph{Runtime Monitoring against Session Types.}

Although the original work on session typing envisaged static checking, a correspondence between MPSTs and communicating automata~\cite{DenielouY12} showed how it was possible to \emph{monitor} processes against a session type, allowing a degree of runtime verification. The key idea is to translate a local type into a finite state machine where transition corresponds to a send or receive action; for each session endpoint, a monitor process observes incoming and outgoing messages --- and upon receiving an invalid message, the monitor drops it and/or reports a violation. We describe runtime monitoring against session types in more depth in Section~\ref{sec:related}.

\paragraph{Multi-language Interoperability.}
There has been increasing attention given to semantic foundations for multi-language interoperability, for example through foreign function interfaces (FFIs). A major inspiration for our proposed solution is the approach of~\citet{PattersonMWA22} who introduce a methodology for interoperability by defining a common intermediate representation along with \emph{convertibility relations} and \emph{boundary conversions} and show how to verify semantic soundness using logical relations.

Our goal is to adopt an analogous methodology but for the world of message passing as opposed to shared memory. Rather than challenges such as linking and foreign function interfaces, our challenge is to describe the ``glue'' that can allow a program written in Go, for example, to safely interact with a program written in Java or Rust, while keeping as many of the guarantees that we would expect if a program was written in a single behaviourally-typed language.

\section{Heterogeneous Multiparty Session Typing}\label{sec:hetero-mpst}

We can start by writing an idealised global type (omitting some irrelevant timing constraints):

{\footnotesize
\[
\begin{array}{l}
\globalsend
    {Customer}
    {Agent}
    {search}
    {\var{origin}: \ty{AirportName}, \var{destination}: \ty{AirportName}} \then \\
\globalsend
    {Agent}
    {Customer}
    {results}
    {\var{searchResults}: [(\ty{FlightNum}, \ty{Time}, \ty{Price}, \ty{Provider})]} \then \\
\globalbranch{Customer}{Agent} \\
\quad \globalmsgann{select}{t \le 360}{\var{flightNum}: \ty{FlightNum}} \mapsto  \\
\qquad \globalsend{Agent}{Customer}{providerRef}{\var{ref}: \ty{ProviderRef}} \then \\
\qquad \globalsendann{Customer}{Provider}{book}{\var{token} == \var{ref}}{\var{token}: \ty{ProviderRef}, \var{details}: \ty{PaymentDetails}} \then \\
\qquad \globalbranch{Provider}{Customer} \\
\qquad \quad \globalmsg{ok}{} \mapsto \globalend, \\
\qquad \quad \globalmsg{error}{} \mapsto \globalend \\
\qquad \}, \\
\quad \globalmsgann{timeout}{t > 360}{} \mapsto \\
\qquad \globalsend{Customer}{Provider}{timeout}{} \then \globalend \\
\}
\end{array}
\]
}

Note that the clock at the customer can only send a $\mkwd{select}$ message within 360 seconds, and otherwise must send a timeout message. Similarly, the data refinement on the $\mkwd{book}$ message ensures that the \emph{same} reference is sent to the provider as is received from the agent.
Say that our system consisted of:

\begin{itemize}
\item The $\role{Agent}$ in Python using the time-aware framework proposed by~\citet{NeykovaBY17}
\item The $\role{Customer}$ in F$\star$ using \textsc{Session}$\star$ by \citet{ZhousFHNY20} to statically verify the data refinement
\item The $\role{Provider}$, developed by a different company and accessible only through an API
\end{itemize}

In the above, the $\role{Agent}$ has inbuilt verification of timing constraints, the $\role{Customer}$ has static verification of data refinements, and the $\role{Provider}$ does not even have verification of communication patterns.
There are several research problems posed by this scenario:

\begin{description}
    \item[Extensibility.]
    At present there are different, incompatible, \emph{dialects} of session types for each extension.
    \item[Precision mismatch.]
    However, each tool available to implement a constraint (Python for timing
    constraints; \textsc{Session}$\star$ for data refinement) only works in a single language. Thus,
    some checks must take place at runtime using boundary monitors.
    \item[Message rejection.] 
    Existing work on monitoring against multiparty session types takes a \emph{suppression-based} approach to monitoring: a monitor will drop any non-conforming messages that it receives (either silently, or reporting a violation). Although suppression maintains safety (by stopping any non-conforming messages from being processed by the program logic), dropping a message may mean that the remaining actions in the protocol cannot be fulfilled---thus breaking liveness.
\end{description}

As well as research questions, there are also some more practical issues:%, requiring engineering and standardisation:

\begin{description}
    \item[Session Initiation.] The session needs to be \emph{established}, which involves discovering each component, inviting it to the session, and setting up the communication infrastructure.
    \item[Wire format.] Communication needs to occur over a \emph{standard message layout}. Most session typing systems either do not include any wire communication, or communicate using non-standard message layouts that vary per tool.
\end{description}

\section{Proposed Solution}\label{sec:solution}

\begin{figure}[t]
\centering
\includegraphics[width=0.77\textwidth]{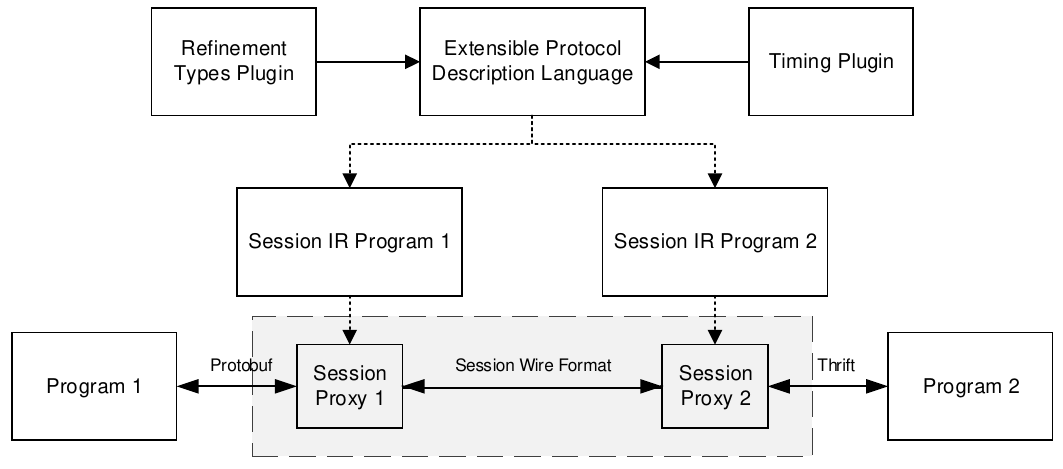}
\caption{Proposed System}
\label{fig:sys-diag}
\end{figure}

Figure~\ref{fig:sys-diag} gives an overview of our proposed approach. The overall idea is:

\begin{itemize}
\item A developer designs the protocol in an \emph{extensible, language-agnostic protocol description language}
\item The protocol is projected and compiled down into a program-specific \emph{session IR}, whose purpose is to describe any necessary dynamic checks, message re-orderings, and data transformations
\item The session IR is used to generate \emph{session proxies} that act as adapters to each program
\end{itemize}

\paragraph*{Extensible Protocol Description Language}
The first step is to write a protocol in a language-agnostic protocol description language. The \emph{Scribble} protocol description language~\cite{YoshidaHNN13, YoshidaZF21} provides a good starting point, but the key point is that the language should be extensible in a modular way by supporting \emph{plugins} supporting individual language features (e.g., value-dependency or timeouts).

\paragraph*{Session IR}
In traditional multiparty session programming, a global protocol description is
projected as a \emph{local type} for each participant. The local type serves as
a language-agnostic \emph{specification} for the communication actions that the
participants should perform, and can be used for static type checking.

In addition to local types for static checking, we propose a \emph{session
intermediate representation} (Session IR).
Whereas local types are meant to be implementation-agnostic, the Session IR is
used to describe any monitoring or message transformations required in order to
integrate a component with the system. 
In particular, we envisage the session IR being able to support at least the
following:

\begin{itemize}
   \item \textbf{Boundary Monitoring} To address the precision mismatches
       between the static checking supported by the tool, as well as maintaining
       timing properties, a core role of each session proxy is to perform
       boundary monitoring~\cite{BocchiCDHY17,ChenBDHY11} before an incoming
       message is delivered (and in some circumstances before an outgoing
       message is committed to the system). Monitor violations may result in
       \emph{suppression} (i.e., dropping messages that are violated), but may
       also allow violations to be reported to the program in order to allow
       compensatory actions (e.g., raising an exception or requesting that the
       sender provides a revised message).
   \item \textbf{Message Insertion / Reordering}
   It may be that a component supports a compatible variation of its role in a protocol, but does not match the protocol exactly (for example, due to a version upgrade). In this case, the session IR can describe message reorderings (inspired by theoretical work on session type isomorphisms~\cite{DezaniPP14,AltayevaY19}), or insertions of messages at a given point. This would enable migrating components to new versions with a different but compatible behaviour.
   \item \textbf{Wire Formatting} Each program will have its own expected communication mechanism (be that sockets, protocol buffers~\cite{protobuf}, or interface description languages like Thrift~\cite{thrift}). The final job of the session IR will be to describe transformations between the internal protocol representation and the wire protocol.
\end{itemize}

For trusted components that are statically checked to follow the protocol, the session proxy will only need to perform wire formatting and monitoring of incoming messages.

\paragraph*{Session Proxies.}
Each session IR program can then be compiled into a \emph{session proxy} to
interact with each program. The session proxy acts as an adapter between the
program and the other participants in the session. Interaction between session
proxies happens using a standardised session wire format and communication
medium.

There are several ways by which session proxies could be integrated with each
application. We expect that for basic suppression-based monitors, it would
suffice to leave the original application untouched and instead route all
communication through the proxy (indeed, this would also support a level of
session typing for components that are not programmed with session types). For
more involved session proxies such as those that raise an application-level
exception when a message violation occurs, it would likely be necessary to
either adapt the current tooling to incorporate the session proxy or provide an
API that a developer can code against.

\paragraph*{Language Features.}
Another concern is the features that must be available to each language or tool in order for it to support any operations required by each session proxy. For suppression-based monitors, it is unlikely that any additional language features would be needed. If we would like to report any monitor violations to the program, however, then we would likely need some additional language support: for example, exception handling~\cite{FowlerLMD19} in the case of needing to deal with linear resources, or additional failure handling callbacks in the case of a tool based around inversion-of-control~\cite{ZhousFHNY20}.

\paragraph*{Session Establishment.}
Figure~\ref{fig:sys-diag} does not describe how sessions are established. We envisage a middleware application, similar to that described by~\citet{Atzei17}, is a potential solution: such a system would allow participants to register to take part in a session, discover other participants, and finally allow sessions to be established. Alternatively, session establishment could happen directly (as is done, for example, with explicit connection actions~\cite{HuY17}).

\subsection{Potential Challenges}

Although we believe our proposed solution offers a promising framework for future research on heterogeneous session typing, we envisage several challenges, at least including the following:

\begin{description}
    \item[Interactions between extensions.]
    Our scenario considers two fairly orthogonal extensions: data refinements and timing. However, there could be other extensions (e.g., explicit connection actions~\cite{HuY17} that require a more liberal syntax) that could pose challenges when combined with existing disciplines. How can we ensure that the extensible language is sufficiently general to both mediate between the different dialects of session typing, and how can we ensure that their combination does not lead to safety errors?
    \item[Formal guarantees.]
    It is important to understand the desired guarantees to be given by the
    system. It seems reasonable to expect, at a minimum, that the system will
    ensure session fidelity (i.e., that every message that is exchanged will
    conform to the given protocol).
    Nevertheless, ensuring properties such as liveness is more challenging in a
    monitored setting.
    A further open challenge would be reasoning about the metatheory in a
    modular way.
    \item[Generality of the IR.]
    The Session IR will at least need to be able to describe monitoring, message reordering, and message reformatting. However, there is a large design space and there are inherent trade-offs to ensuring the IR design remains sufficiently high-level while also sufficiently general to support the array of possible extensions.
    \item[Performance.]
    Any runtime checking and message rewriting will inevitably incur a runtime
    overhead, so it will be necessary to ensure that any overhead is not
    prohibitive.
    This can be mitigated to an extent by only checking properties that are not
    guaranteed statically, and ensuring that the monitor is located on the same
    machine as the monitored process.
    In addition to runtime overheads, it is possible that monitors may need to record some message history, for example to enforce dependent type constraints~\cite{ToninhoY17}. It is therefore necessary to ensure that any generated monitor does not require unbounded space.
    \item[Location of error reporting.]
    In addition to the language design challenges in allowing applications to handle any errors,
    it is possible that there are multiple places to report a violation. Some notion of blame~\cite{WadlerF09} is likely to be important, but defining ``more'' or ``less'' typed is likely to be challenging in the presence of multiple extensions.
\end{description}

\section{Related Work}\label{sec:related}

\paragraph*{Protocol description languages.}
MPSTs were designed without a particular implementation in mind. A good starting point for heterogeneity is the language-agnostic Scribble protocol description language~\cite{YoshidaHNN13,NeykovaY19} for describing MPST specifications; implementations of Scribble~(e.g.,~\cite{YoshidaZF21}) support well-formedness checking, projection, and monitor generation.  A necessary step in the pursuit of heterogeneity would be to generalise a language like Scribble to allow modular and composable
extensions with different language features, as opposed to the current status quo of \emph{ad-hoc} extensions for each new feature.

\paragraph*{Monitoring against MPSTs.}
Most closely relevant is work on runtime monitoring against local
types~\cite{BocchiCDHY17,ChenBDHY11}, based on the correspondence between
multiparty session types and communicating automata~\cite{DenielouY12}. The core
idea is to translate each local type into an FSM and check each incoming and
outgoing message against the monitor, rejecting the message if the message does
not match any transition. In particular,~\citet{BocchiCDHY17} show \emph{safety}
(monitors ensure that ill-behaved participants do not send messages that violate
their specification) and \emph{transparency} (monitors do not affect the
behaviour of well-behaved components).
However, these monitors discard non-conforming messages without any
feedback to the sender or receiver. In turn, this means that (especially for
non-recursive protocols), non-conforming messages cause the protocol to silently
stop. Later work by \citet{HeuvelPD23} addresses the black-box monitoring of
multiparty sessions through monitoring processes that are directly generated
from global types, and (unlike \citet{BocchiCDHY17}) actively report violations
by stopping execution.
\citet{BurloFS20} proposes a prototype implementation of
a ``hybrid'' verification approach where session-typed components can
interoperate with heterogeneous (possibly untyped) components through
autogenerated monitors --- which, in turn, are session-typed (only for two-party
sessions), and can translate messages between different wire formats, and
suppress and report protocol-violating messages;
\citet{DBLP:conf/ecoop/BurloFS21} later studies the properties of such black-box
monitors in terms of soundness and completeness of violation reports.
In contrast to all the works on session monitoring listed above, we believe that
it may be necessary to forego monitor transparency and instead allow
compensatory behaviours upon a monitor violation (for example, raising an
exception or requiring the sender to send a revised message).

\paragraph*{Session types and heterogeneity.}
Only very little work has attempted to address heterogeneous session typing. \citet{JongmansP22} describe the design of a system called ST4MP that aims to support multi-lingual programming through the established API generation approach~\cite{HuY16}; the idea is to generate multiple compatible APIs for different languages from a given global type specification. In contrast to this position paper, ST4MP supports a base multiparty session typing discipline without any advanced language features (e.g., timing or refinement types), and does not make use of any form of runtime checking. In contrast we would expect our general session IR to be able to support even untyped components.

\paragraph*{Language and system interoperability.} Our proposal to use a common Session IR to provide safety properties even when composing heterogeneous components written in different languages is similar in spirit to recent work on sound language interoperability~\cite{PattersonMWA22, PattersonWA23}. While previous work only targets languages interoperating via shared memory, our proposal specifically aims to address interoperability for typed message-passing concurrency. Gradual session types~\cite{IgarashiTTVW19} provide a framework for ensuring type and communication safety for programs integrating statically-typed sessions and dynamic types. It is assumed that programs share a common internal language with casts. Our proposal aims to decouple components even further by mediating communication via Session IR proxies which may, in addition to casts, insert or reorder messages and perform other forms of monitoring. Session IR and Session IR proxies are related to IDLs used for integrating distributed components~\cite{Lewandowski98} as well as systems for business-to-business interactions~\cite{MedjahedBBNE03} which explicitly aim to address heterogeneity.

At a more abstract level, the ideas presented in this position paper can be related to another position paper by Albert \emph{et al.}~\cite{engineering-virtualized-services} that advocates a formal language for service-level agreements (SLAs) for (virtualised and heterogeneous) distributed services, to be enforced via e.g.~static verification and/or runtime monitoring, possibly aided by code generation from executable models written in ABS~\cite{DBLP:conf/fmco/JohnsenHSSS10}. Our approach is
focused on behavioural types (which could be seen both as a form of formalised SLA) and as a form of executable specification (usable e.g.~for monitor generation via the session IR).

Jolie~\cite{MontesiGZ14, jolie-website} is a service-oriented programming
language. Programs can either be written in Jolie, or Jolie can serve as an
interface to services written in a different programming language. Jolie
programs can also serve as \emph{orchestrators} to interact with multiple other
services, potentially using different transport mechanisms.
In contrast, our proposal takes a more protocol-centric approach: instead of
specifying the services and an orchestration-based method of allowing them to
interact, our proposal instead involves concentrates on boundary monitoring and
manipulation of existing communication flows.
An advantage is that we can (potentially statically) verify fine-grained data
and timing constraints.

\section{Conclusion}\label{sec:conclusion}
Although behavioural types offer a strong foundation for lightweight, language-integrated verification of behavioural properties, a large barrier to their adoption is that at present an \emph{entire system} must be written in a single language.
In this position paper we have described a potential line of work that, if
completed successfully, could allow behavioural types in \emph{heterogeneous}
software systems where components can be written in different languages, using
different tools, each of which support different static guarantees. Our approach
relies on an  \emph{extensible} protocol description language that can support
additional language features (e.g., timing or refinement types) as plugins, and
a \emph{session IR} that can describe transformations on data (e.g., wire formatting, message reordering, and boundary monitoring).
Structured support for heterogeneity can greatly expand the reach of behavioural types in real-world systems, and we hope that these initial ideas serve as a starting point for addressing this challenging research topic.

{\footnotesize
\section*{Acknowledgements}
We thank the organisers of Dagstuhl Seminar 24051 and Schloss Dagstuhl --- Leibniz Center for Informatics for making this work possible.
We are also grateful to the anonymous reviewers for their detailed and encouraging reviews.
This work was supported by EPSRC grant EP/T014628/1 (STARDUST), Horizon Europe
grant 101093006 (TaRDIS), Independent Research Fund Denmark RP-1 grant
``Hyben'', Digital Futures Research Pairs Consolidator Project ``PORTALS'', and
UKRI Future Leaders Fellowship MR/T043830/1 (EHOP), FCT grant PTDC/CCI-COM/6453/2020 (SafeSessions), and the LASIGE Research Unit.
}

%%\nocite{*}
\bibliographystyle{plainnat}
\bibliography{heterogeneous}
\end{document}